# The impact of air transport availability on research collaboration: A case study of four universities

Adam Ploszaj[1*], Xiaoran Yan[2], Katy Börner[2,3]

[1] Centre for European Regional and Local Studies EUROREG, University of Warsaw, Warsaw, Poland
[2] Indiana Network Science Institute, Indiana University, Bloomington, Indiana, United States of America
[3] School of Informatics, Computing, and Engineering, Indiana University, Bloomington, Indiana, United States of America

* Corresponding author
E-mail: a.ploszja@uw.edu.pl (AP) (ORCID: 0000-0002-6638-3951)

## ABSTRACT

This paper analyzes the impact of air transport connectivity and accessibility on scientific collaboration. Numerous studies demonstrated that the likelihood of collaboration declines with increase in distance between potential collaborators. These works commonly use simple measures of physical distance rather than actual flight capacity and frequency. Our study addresses this limitation by focusing on the relationship between flight availability and the number of scientific co-publications. Furthermore, we distinguish two components of flight availability: (1) direct and indirect air connections between airports; and (2) distance to the nearest airport from cities and towns where authors of scientific articles have their professional affiliations. We provide evidence using Zero-inflated Negative Binomial Regression that greater flight availability is associated with more frequent scientific collaboration. More flight connections (connectivity) and proximity of airport (accessibility) increase the number of coauthored scientific papers. Moreover, direct flights and flights with one transfer are more valuable for intensifying scientific cooperation than travels involving more connecting flights. Further, analysis of four organizational sub-datasets—Arizona State University, Indiana University Bloomington, Indiana University-Purdue University Indianapolis, and University of Michigan—shows that the relationship between airline transport availability and scientific collaboration is not uniform, but is associated with the research profile of an institution and the characteristics of the airport that serves this institution.

# INTRODUCTION AND PRIOR WORK

Despite the proclaimed "death of distance" (Cairncross, 1997; Friedman, 2005), geography is of constant importance for scientific collaboration (Morgan, 2004; Olson & Olson, 2000; Olechnicka et al., 2018). Numerous studies demonstrated that the likelihood of collaboration declines with growing distance between prospective collaborators. This effect is observed both at the micro level of buildings or campuses, as well as at the macro level of collaboration networks among cities, regions, and countries.

At the micro level, Allen (1997) showed in the 1970s that the frequency of communication between individuals in science and engineering organizations drops exponentially with the growing distance between their offices. Subsequent research revealed that collaboration is more likely not only between closely sited or collocated individuals (Boudreau et al., 2017; Catalini, 2018) but also between those whose daily paths cross frequently or largely overlap (Kabo et al., 2014; Kabo et al., 2015).

At the macro level—where distance is measured in kilometers rather than meters—a large body of evidence indicates the negative impact of spatial separation on research collaboration: the greater the distance, the lower the likelihood of collaboration. Furthermore, geographical distance not only decreases the likelihood of any collaboration, but also reduces the intensity of collaboration, as measured by the number of co-publications, co-patents, and collaborative projects (Adams, 2013; Fernández et al., 2016; Katz, 1994). The relationship between distance and collaboration is frequently analyzed in the framework of the general gravity model (Hua & Porell, 1979). The gravity model is conceptually based on Isaac Newton's law of gravitation. It says that the gravitational force between two objects is proportional to their masses and inversely proportional to the square of the distance between them. The model assumes that not only the distance between collaborating units matters, but also their "masses" should be taken into account. Here "mass" refers to research capacity of the collaborating units, typically measured by research and development employment or expenditures, as well as by accumulated research outputs: stocks of funded projects, publications, and patents. The gravity model applied to scientific collaboration clearly shows that the probability and intensity of research collaboration are negatively related to the geographic distance which separates the units in question and are positively affected by their accumulated research potential (Andersson & Persson, 1993; Hoekman et al., 2009; Hoekman et al., 2010; Hoekman et al., 2013; Picci, 2019; Plotnikova & Rake, 2014; Sebestyén & Varga, 2013).

The detrimental effect of geographical distance on the likelihood of research collaboration remains significant even when controlling for important features of collaborating units, type of collaborative relations, and the context in which collaboration occurs. Previous studies controlled for scientific quality, most frequently measured via citations (Bianconi & Barabási, 2001; Ke, 2013; Mazloumian et al., 2013), differences in cooperation patterns accross various fields of science (Barber & Scherngell, 2013; Franceschet & Costantini, 2010; Gingras, 2016; Larivière et al., 2006), type of research (Wagner, 2008), and the type of collaboration data used in the analysis, such as co-publications, co-patents, and collaborative projects (Lata et al., 2015; Zitt et al., 2003). Prior work has also considered different types of non-spatial proximities, including cognitive, cultural, economic, institutional, organizational, social, and technological (Boschma, 2005; Capello & Caragliu, 2018; Knoben & Oerlemans, 2006; Nagpaul 2003; Marek et al., 2017).

The rise in research collaboration manifests itself not only in the growing number of co-authors per paper (and co-inventors per patent), but also in the increasing co-authorship among authors whose institutional affiliations were in different countries. Between 1990 and 2011, the percentage of internationally co-authored papers indexed in the Science Citation Index increased from 10.1% to 24.6% (Wagner et al., 2015). Co-authorship is particularly intense between authors affiliated with the largest research centers, which serve as major hubs in the global scientific cooperation network (Matthiessen et al., 2010; Maisonobe et al., 2016). At the same time, researchers are increasingly collaborating across greater distances. Between 1980 and 2009 the mean collaboration distance per publication raised from 334 to 1,553 kilometers (Waltman et al., 2011).

The distance between collaborating units in spatial scientometrics studies is usually measured as geographical distance along the surface of the earth ("as the crow flies"), between points which are defined by geographical coordinates: latitude and longitude (Frenken et al., 2009). The actual accessibility is taken into account surprisingly rarely in empirical studies of scientific collaboration. To our best knowledge, only following empirical works considered actual transport accessibility as a covariate of scientific collaboration. Andersson and Ejermo (2005) included road travel time in their case study of Swedish patent co-authorship network. Ejermo and Karlsson (2006) studied road and air travel time impact on co-patenting in Sweden. Ma, Fang, Pang, and Li (2014) hypothesized that high-speed railway accessibility can be one of the factors explaining the intensity of scientific cooperation between Chinese cities. Later, the hypothesis was supported with evidence from instrumental variable regression study designed by Dong, Zheng, and Kahn (2018). Furthermore, Hoekman, Frenken, and Tijssen (2010) argued that European regions with a major international airport are more likely to develop intensive international scientific collaboration. Against this background, the study of Catalini, Fons-Rosen, and Gaulé (2016) stands out as the authors used a quasi-experimental design (natural experiment) to examine the impact of introducing a new, low fare, air route on the probability of scientific cooperation. Their analysis focuses on 890 faculty members in chemistry departments of research-intensive US universities in the period from 1991 to 2012. The results show that the introduction of new routes significantly increases the likelihood of collaboration among US chemistry scholars. The greatest impact is observed in the case of early career scholars, who usually have fewer resources than established professors do, and therefore cheaper flights may be more important to them.

Our study extends prior work by analyzing the relationship between scientific collaboration and worldwide air transport availability. We distinguish two components of flight availability: (1) direct and indirect air connections between airports (connectivity), and (2) distance to the nearest airport (accessibility) from cities and towns where scientific articles are affiliated. We test the hypothesis that better air transport connectivity and accessibility—ceteris paribus—is positively associated with scientific collaboration. Furthermore, we hypothesize that the relation depends on research capacity and profile of a given university and the flight network of an airport that serves the university. To account for such specific circumstances, we selected four campuses of US public research-intensive universities: Arizona State University at Tempe (ASU), Indiana University Bloomington (IUB), Indiana University-Purdue University Indianapolis (IUPUI) and University of Michigan at Ann Arbor (UMICH). Only the main campuses of the universities are included in the study. Our selection criteria comprised comparable size and research intensity of universities, various levels of passenger traffic, and the possibility of an unambiguous assignment of a major research university to a single airport. ASU is served by Phoenix Sky Harbor International Airport (PHX) and UMICH by Detroit Metropolitan Airport (DTW). Both airports are important hubs. According to Federal Aviation Administration data, PHX was the 11th US airport in terms of number of passengers in 2016, while DTW took 18th position. IUB and IUPUI constitute a specific case. The two campuses are served by the same airport, Indianapolis International Airport (IND). IND is an airport with considerably less passenger traffic than PHX and DTW. In 2016, it was 46th US airport regarding the number of passengers.

The remainder of the paper is organized as follows. The next section introduces our empirical strategy and presents variables and descriptive statistics. Then, we present our approach to model the relation between the number of co-authored papers and air transport availability. We then discuss findings. The paper concludes with discussion and conclusions. Supporting information includes detailed information on data sources and data processing procedures, as well as information needed to replicate the results of this study.

# EMPIRICAL STRATEGY AND DESCRIPTIVE STATISTICS

The number of co-authored papers is the dependent variable in this study. Co-authorship were identified on the basis of the co-occurrence of author affiliations in articles published in years 2008-2013 and indexed in the Web of Science database. We employed the full counting method, i.e. each co-authored paper is counted as one for a given ego-alter relation, regardless of the number of authors, organizations, geo-locations or countries involved (Perianes-Rodriguez at al., 2016). The advantage of this approach— as compared to fractional counting—is the intuitive interpretation of results, as well as the possibility of using well-established statistical models for event counts data (Long, 1997).

The dependent variable is measured for each of four institutions—ASU, IUB, IUPUI, and UMICH—as the number of co-authored papers between the given campus and various geographical units across the globe (henceforth called as 'destinations'). To ensure coherence and international comparability geo-locations are merged into 2,245 town/city/metropolitan/regional entities, such as European NUTS2 regions and US Metropolitan Statistical Areas (see Fig 1). For each of four selected universities a separate egocentric co-authorship network was constructed. In consequence, we obtained four ego-networks, in which an ego was ASU, IUB, IUPUI or UMICH, and alters (destinations) were spatial units from around the world (for the details on data sources and data processing, please refer to the Supplementing information[1]).

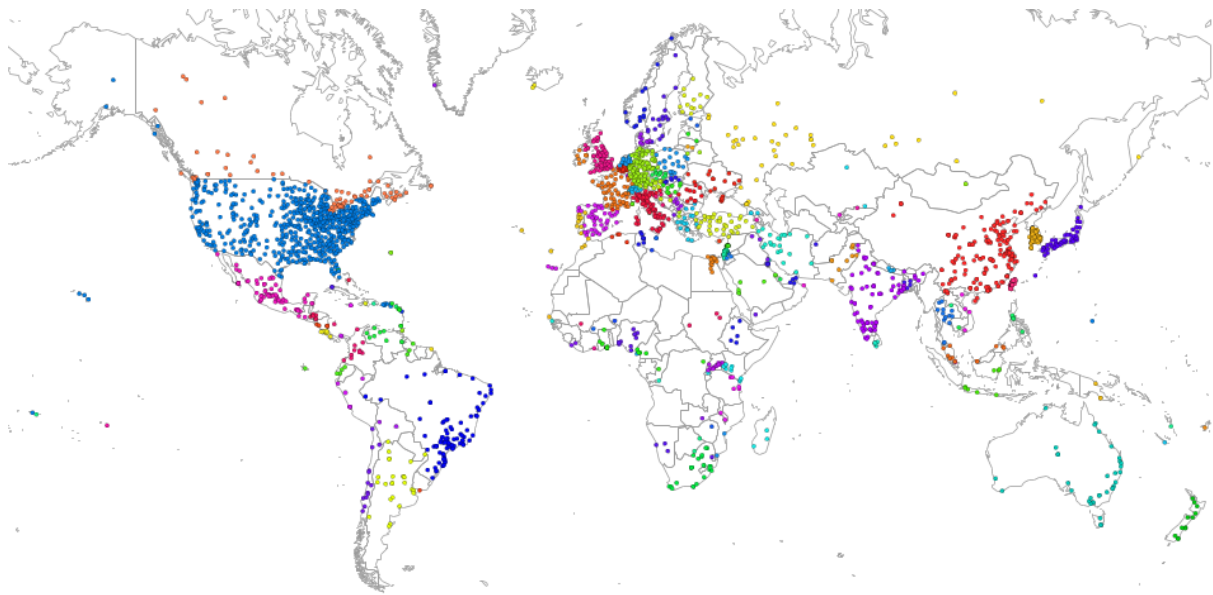

**Fig 1. Merged cities and metro areas under this study**
* Colours represent countries.

To measure air transport availability we employed a number of variables grouped into two categories: commercial air transport connectivity and transport accessibility to the nearest airport. The accessibility variable is measured as the geographical distance from the center (centroid) of a destination to its nearest airport with commercial flights. To account for connectivity, we tested three approaches. The most straightforward variable is a 'Minimum number of stops to reach destination'. This factor variable is based on a minimum number of connecting flights needed to travel from ego's nearest airport to the airport nearest to the centroid of destination geographical unit. It is measured up to 4 connecting flights (or 3 stops) and takes values: 0 (for direct flights), 1, 2, or 3. Second measure 'LinesXstop' takes into account number of flights between ego and destination airports. 'Lines0stop' accounts for direct flights only. 'Lines1stop' measures direct and indirect flights up to one stop (i.e., up to two connecting flights). 'Lines2stop' considers direct and indirect flights up to two stops, while 'Lines3stop' adds connections requiring 3 stops. To take into account the preference for flights with fewer transfers, weights are applied: 1 for direct flights, 0.5 for one stop connections, 0.33 for two stop, and 0.25 for three stops. 'SeatsXstop' variable is constructed in a similar way, but it also takes into account number of seats available on direct and connecting flights. The use of concurrent connectivity variables aims to better understand the relationship between air transport and scientific collaboration. Three questions are particularly interesting in this case. First, are direct connections more important than connecting flights? Second, are indirect flights with fewer stops more important than those with more stops? Third, does the passenger capacity (number of available seats) matter?

---

[1] Available on GitHub: https://github.com/everyxs/FlightCoauthor

Two control variables are used in this study. 'Geographical distance' between an ego-institution and a destination is measured along the surface of the earth. We assume that geographical distance alone should explain a lot of scientific collaborations. However, we hypothesize that models accounting simultaneously for geographical distance and flights availability variables will fit the data better. The second control variable is the 'Number of papers at destination'. This variable can be seen as the equivalent of a mass term in the gravity model approach. We assume that probability and intensity of collaboration between ego and destination depend primarily on the scientific capacity of a destination. Collaboration with city, region, or country that have virtually no research activities is improbable. While collaboration with global knowledge hubs, e.g. Oxford, Paris, or Tokyo, can be intensive, despite the geographical distance.

Our full dataset of 8,980 observations (units of analysis) consists of four institutional sub-datasets, each comprising 2,245 observations (see Table 1 and 2). An observation is defined as a multidimensional link (co-authorships, geographical distance, air links, etc.) between university campus in question—one of the four ego-institutions—and one of 2,245 geographical entities around the world that have at least one paper affiliated as identified by Mazloumian et al. (2013). The number of co-authored papers between ego-institution and defined geographical entities—the dependent variable in this study—ranges from 0 to 3433, with the mean value of the variable equal to 15.4 (in the period of 2008-2013). It means that the four analyzed institutions co-authored on average 15.4 papers per possible relationship between the institution and one of the defined geographical units. In this regard, UMICH stands out from the other three universities. Its average number of papers co-authored with researchers affiliated with institutions located in other spatial units around the world equals 34.6, while for other institutions it lays in the range from 8.2 to 9.9.

**Table 1. Descriptive statistics – full dataset**

| Variable | Observations | Mean | Std. Dev. | Min | Max |
|---|---|---|---|---|---|
| Number of co-authored papers | 8980 | 15.4 | 89.5 | 0 | 3433 |
| Geographical distance (mi) | 8980 | 4232.3 | 2669.4 | 20.4 | 11171 |
| Number of papers at destination | 8980 | 5373.3 | 13866 | 1 | 201693 |
| Distance to airport at destination (mi) | 8980 | 24.8 | 25.4 | 0.4 | 327 |
| lines0stop | 8980 | 0.1 | 0.7 | 0 | 15 |
| lines1stop | 8980 | 3.8 | 6 | 0 | 55 |
| lines2stop | 8980 | 18 | 16.8 | 0 | 127 |
| lines3stop | 8980 | 114.6 | 91.6 | 0 | 822 |
| seats0stop | 8980 | 24.1 | 128.5 | 0 | 2016 |
| seats1stop | 8980 | 623 | 1049.3 | 0 | 8523 |
| seats2stop | 8980 | 3071.3 | 3002.3 | 0 | 21249 |
| seats3stop | 8980 | 95361.2 | 153682.8 | 0 | 1535855 |
| Min. number of stops to destination | 8980 | 1.5 | 0.7 | 0 | 4 |

The geographical distance between the four ego-institutions and their collaborators varies from 20.4 to 11,171 miles. Mean geographical distance between all possible dyads (between one of the four ego-institutions and all other possible collaborators in their network) is 4,232 miles. To put this number in context, recall that the distance between New York City and Los Angeles is about 2,450 miles. The high average geographical distance results for the fact that many coauthors have institutional homes on other continents. UMICH has the lowest mean geographical distance between it and collaborating institutions (4,001 miles), followed by UIPUI and IUB (4,080 and 4,085 miles respectively), while ASU is characterized by the highest geographical separation from its collaborators (4,232 miles). The juxtaposition of the number of co-authored papers and the distance between co-authors' affiliations reveals that collaboration is not uniformly distributed across geographic space (see Fig 2). A pattern is evident across all four institutions: A university substantial proportion of collaborations take place in the range up to 2,000 miles, there are almost no collaborations in the 2,000 to 4,000 mile range, then, from over 4,000 miles (over 5,000 miles in the case of ASU) collaborations are again evident. Comparing these distances to a map shows that the closest set of collaborations reflects those in which the collaborator is within the continental U.S. or North America, the gap at 2,000 to 4,000 miles reflects the Atlantic and Pacific Oceans, and the range from 4,000 to 6,000 miles reflects mainly U.S.—European collaborations.

**Table 2. Descriptive statistics – institutional sub-datasets**

| Variable | Observations | Mean | Std. Dev. | Min | Max |
|---|---|---|---|---|---|
| **ASU** | | | | | |
| Number of co-authored papers | 2245 | 9.9 | 40.8 | 0 | 793 |
| Geographical distance (mi) | 2245 | 4762.6 | 2619.2 | 82.9 | 10934 |
| Number of papers at destination | 2245 | 5375.3 | 13875.2 | 1 | 201693 |
| Distance to airport at destination (mi) | 2245 | 24.8 | 25.4 | 0.4 | 327 |
| Lines0stop | 2245 | 0.3 | 1 | 0 | 15 |
| Lines1stop | 2245 | 4.4 | 7.3 | 0 | 55 |
| Lines2stop | 2245 | 20.6 | 19.4 | 0 | 127 |
| Lines3stop | 2245 | 131.7 | 104.7 | 0 | 822 |
| Seats0stop | 2245 | 42.4 | 185.7 | 0 | 2016 |
| Seats1stop | 2245 | 759.1 | 1300.9 | 0 | 8523 |
| Seats2stop | 2245 | 3653.5 | 3495.3 | 0 | 21249 |
| Seats3stop | 2245 | 127024.4 | 182298.8 | 0 | 1521228 |
| Min. number of stops to destination | 2245 | 1.4 | 0.7 | 0 | 4 |
| **IUB*** | | | | | |
| Number of co-authored papers | 2245 | 8.2 | 30.6 | 0 | 469 |
| Geographical distance (mi) | 2245 | 4085.3 | 2684.7 | 20.4 | 11075 |
| Number of papers at destination | 2245 | 5380.2 | 13879.6 | 1 | 201693 |
| Distance to airport at destination (mi) | 2245 | 24.8 | 25.4 | 0.4 | 327 |
| Lines0stop | 2245 | 0 | 0.4 | 0 | 9 |
| Lines1stop | 2245 | 2.8 | 4.8 | 0 | 37 |
| Lines2stop | 2245 | 15 | 14.1 | 0 | 95 |
| Lines3stop | 2245 | 95.7 | 76.6 | 0 | 648 |
| Seats0stop | 2245 | 6.3 | 49.7 | 0 | 1115 |
| Seats1stop | 2245 | 419.4 | 737.2 | 0 | 4937 |
| Seats2stop | 2245 | 2375.4 | 2253.6 | 0 | 13831 |
| Seats3stop | 2245 | 60251.8 | 120158.7 | 0 | 1105416 |
| Min. number of stops to destination | 2245 | 1.6 | 0.7 | 0 | 4 |
| **IUPUI*** | | | | | |
| Number of co-authored papers | 2245 | 9.1 | 44.6 | 0 | 822 |
| Geographical distance (mi) | 2245 | 4080.4 | 2683.5 | 40.5 | 11095 |
| Number of papers at destination | 2245 | 5375.8 | 13875.8 | 1 | 201693 |
| Distance to airport at destination (mi) | 2245 | 24.8 | 25.4 | 0.4 | 327 |
| Lines0stop | 2245 | 0 | 0.4 | 0 | 9 |
| Lines1stop | 2245 | 2.8 | 4.8 | 0 | 37 |
| Lines2stop | 2245 | 15 | 14.1 | 0 | 95 |
| Lines3stop | 2245 | 95.7 | 76.6 | 0 | 648 |
| Seats0stop | 2245 | 6.3 | 49.7 | 0 | 1115 |
| Seats1stop | 2245 | 419.4 | 737.2 | 0 | 4937 |
| Seats2stop | 2245 | 2375.4 | 2253.6 | 0 | 13831 |
| Seats3stop | 2245 | 60251.8 | 120158.7 | 0 | 1105416 |
| Min. number of stops to destination | 2245 | 1.6 | 0.7 | 0 | 4 |
| **UMICH** | | | | | |
| Number of co-authored papers | 2245 | 34.6 | 164.2 | 0 | 3433 |
| Geographical distance (mi) | 2245 | 4000.7 | 2619.9 | 30.4 | 11171 |
| Number of papers at destination | 2245 | 5361.7 | 13842.6 | 1 | 201693 |
| Distance to airport at destination (mi) | 2245 | 24.8 | 25.4 | 0.4 | 327 |
| Lines0stop | 2245 | 0.2 | 0.8 | 0 | 9 |
| Lines1stop | 2245 | 5 | 6.5 | 0 | 50 |
| Lines2stop | 2245 | 21.4 | 18 | 0 | 122 |
| Lines3stop | 2245 | 135.2 | 97.1 | 0 | 805 |
| Seats0stop | 2245 | 41.3 | 159.4 | 0 | 1295 |
| Seats1stop | 2245 | 894.2 | 1204.7 | 0 | 8396 |
| Seats2stop | 2245 | 3880.8 | 3424.7 | 0 | 21114 |
| Seats3stop | 2245 | 133916.5 | 165648.9 | 0 | 1535855 |
| Min. number of stops to destination | 2245 | 1.2 | 0.7 | 0 | 4 |

* IUB and IUPUI are served by one airport, Indianapolis International Airport (IND), therefore they have the same values of air transport variables.

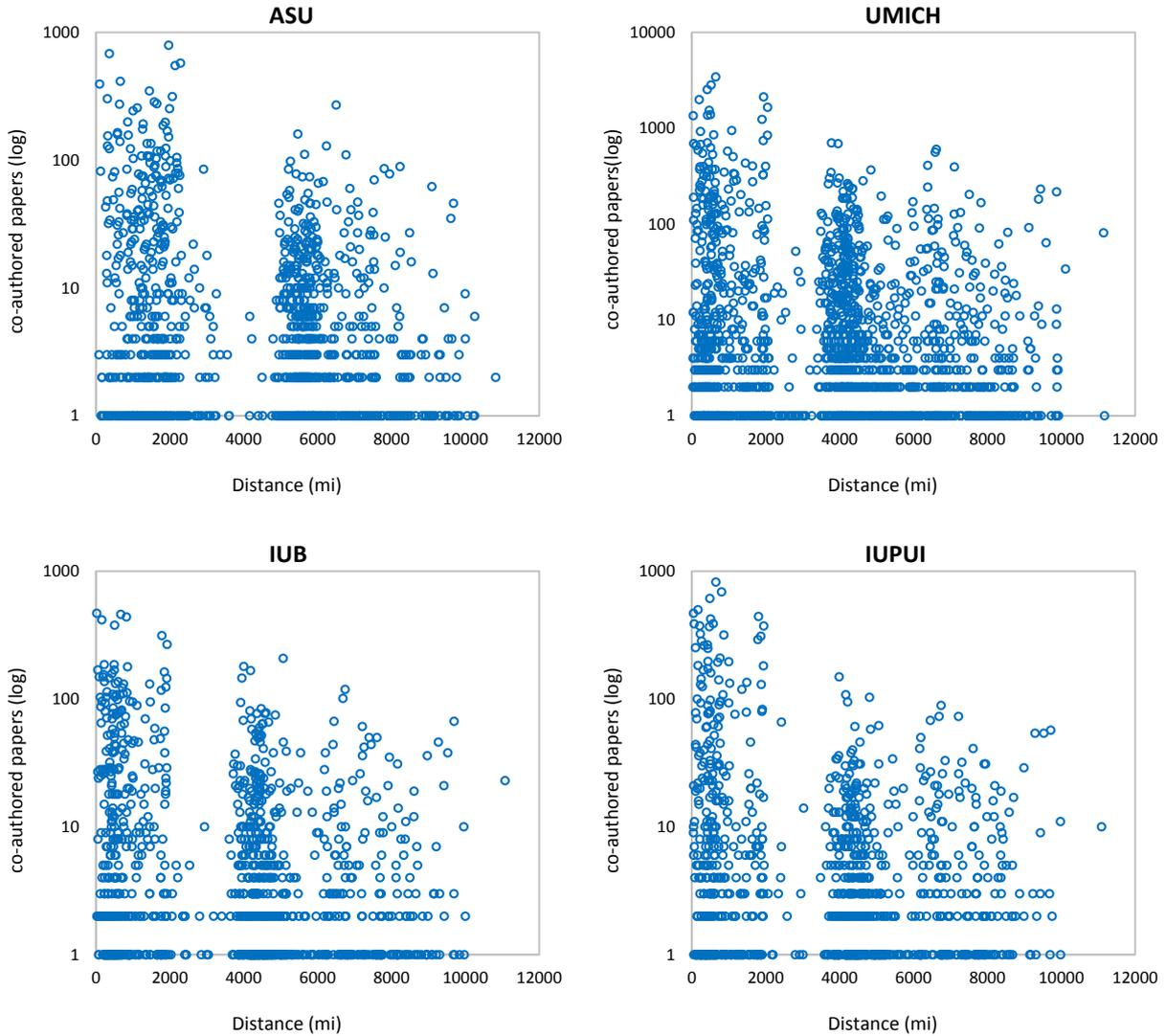

**Fig 2. Co-authored papers distribution by geographic distance**

Descriptive statistics of 'Number of papers at destination' and 'Distance to airport at destination' are almost identical for the full dataset and each of institutional datasets. This is due to the fact that each university has the same set of possible collaborators, except for itself—i.e., ASU ego network excludes ASU, IUB ego network excludes IUB, etc. The number of papers at destination was as low as one (recall that only geographical entitles with at least one affiliated paper were included in the dataset), and as high as almost 202 thousand (Boston metropolitan area). The mean distance from collaborating destination to its nearest airport was about 25 miles. The longest distances to the nearest airport with scheduled flights occur in vast and sparsely populated countries, such as Russia or Canada, and in emerging economies, mainly in Africa and South America.

The values of air transport connectivity variables vary substantially among the four institutional sub-datasets. Three airports that serve four considered campuses—note that IUB and IUPUI are served by a single airport, IND, located on the outskirts of Indianapolis—differ regarding the number of direct flights to collaborative destinations. Consequently, they also differ in the number of collaborating destinations that reachable by direct flights, as well as flights with one, two or three stopovers/connections. UMICH is served by Detroit Metropolitan Airport (DTW) and has a privileged position owing to the fact that scholars from Ann Arbor can reach collaborators in 301 different collaborating destinations via direct flights. Phoenix Sky Harbor International Airport (PHX) serves ASU and provides direct connections to 218 destinations, whereas IND airport only provide 53 direct-flight-accessible destinations. Furthermore, UMICH scholars can travel to more destinations using one-stop connecting flights than scholars from other three universities. On the other hand, for ASU, IUB and IUPUI researchers more destinations are available only via connecting flights with at least two stops (see Table 3). As a result, air transport connectivity variables—'LinesXstop', 'SeatsXstop', and 'Minimum number of stops to destination'—have higher values for UMICH, than in the case of ASU and, in particular, IUB and IUPUI (see Table 2).

**Table 3. Destinations reachable with direct and connecting flights from airports serving four studied universities**

| Airport(s)   | Direct | 1 stop | 2 stops | 3 stops | Total |
|--------------|--------|--------|---------|---------|-------|
| Detroit      | 301    | 1255   | 658     | 31      | 2245  |
| Indianapolis | 53     | 894    | 1134    | 164     | 2245  |
| Phoenix      | 218    | 913    | 1042    | 72      | 2245  |

Detroit serves UMICH, Indianapolis serves IUPUI and IUB, while ASU is served by Phoenix.

# MODELING APPROACH

To model the impact of air transport availability on scientific collaboration we employed zero-inflated model. This class of models is designed for event count data where the sample is drawn from a zero-inflated probability distribution—i.e., one that allows for frequent zero-valued observations. Our research dataset fits the requirements for using these models perfectly—about 45% of the outcome variable equals zero. That is, during the observed period, the four ego-institutions had no co-authorships with 45% localizations that are identified as having published at least one scientific paper (according to data from Mazloumian et al., 2013). The zero-inflated model assumes that zero outcome can result from two different processes. First, the absence of collaboration can be due to the lack of research capacities at the destination. In this case, the expected outcome is zero. Second, if the destination has some research capacities, it is then a count process. Zero outcome is still possible (e.g. due to different research profiles), but numerous co-authorships are very likely.

Consequently, the zero-inflated model has two components: "inflate" part that accounts for excess zeros (the equivalent of logit model) and a proper "count" part. To construct inflate part we used a single predictor: 'Number of papers at destination'. This decision is based on the assumption that the adequate critical mass of scientific capacity determines the emergence of scientific collaboration, regardless of geographical distance and transport accessibility. In the count part, we used both control variables—i.e.

'Geographical distance' and 'Number of papers at destination'—and independent variables for air transport connectivity and accessibility.

To account for expected curvilinearity, additional quadratic terms have been used in the case of four variables: 'Geographical distance', 'Number of papers at destination', 'LinesXstop', and 'SeatsXstop'. We assume that the impact of enumerated variables on scientific collaboration is not uniformed across their possible values. In particular, the impact can be more pronounced at low values and gradually less distinct at high values (diminishing returns pattern). For example, we can expect that the difference between one and two direct flights between the same two cities should have substantial impact on the likelihood of research collaboration, while the difference between 11 and 12 direct flights can have less pronounced effect.

Because air transport makes little sense for short distances, observations in which geodistance variable was less than 100 miles were excluded from the further empirical analysis. In total, 55 observations were omitted, of which 4 for ASU, 23 for IUB, 22 for IUPUI, and 24 for UMICH. As a result, a restricted dataset used as a basis for estimations consisted of 8,925 observations, multidimensional links (co-authorships, geographical distance, air links, etc.) links between four universities and theirs possible research collaborators. Sub-datasets for individual universities were as follows: ASU—2,241 observations, IUB—2,222, IUPUI—2,223, and UMICH—2,221.

We used Zero-inflated Negative Binomial Regression (ZINBR) model implemented in STATA. However, we tested other models for count data: Poisson (PRM), Zero-Inflated Poisson (ZIP), and Negative Binomial Regression Model (NBRM). The results of estimation strongly suggest that ZINBR fits our data significantly better than PRM, ZIP, and NBRM.

The results section of the paper presents model specifications grouped into four tables. Specifications differ in terms of employed independent variables, as well as observations taken into account. Models from (1) to (14) are based on the full dataset, while models (15)-(34) are based on institutional sub-datasets. Model (1) is a reference model that includes only control variables and any of the air transport variables. Other models include various configurations of air transport accessibility and connectivity variables. The comparison of complete and restricted specifications allows for insights into complex relationships between scientific collaboration, air transportation, and geographic separation.

## RESULTS

Table 4 presents estimation results of models with air transport connectivity and accessibility (models 6-9), as well as models without airport accessibility variable (2)-(5), compared to the reference model that does not include any transport variables (1). As expected, the basic model (1) with no air transport availability variables does significantly worse than all other models with transport variables included. This is evidenced by the fact that model (1) has the highest values of Akaike Information Criterion (AIC) and Bayesian information criterion (BIC). The difference in AIC and BIC between the model (1) and the second worst specification, model (2), highly exceeds 10 and can, therefore, be considered significant (Burnham & Anderson, 2002; Raftery, 1995). The addition of air connectivity variables (models 2-5) noticeably improves the fit of the model (significant decrease in both AIC and BIC). Moreover, enriching the model with a variable describing the accessibility of the nearest airport (models 6-9) improves the fit even more. Consequently, models combining air transport connectivity and accessibility (6)-(9) perform significantly better than specifications comprising only connectivity variables (1)-(5). These results plainly indicate that not only the physical distance influences the intensity of scientific collaboration, but also, the actual transport accessibility plays a significant role.

**Table 4. Research collaboration and air transport connectivity and accessibility**

| Dependent variable: Number of co-authored papers | (1) | (2) | (3) | (4) | (5) | (6) | (7) | (8) | (9) |
|---|---|---|---|---|---|---|---|---|---|
| **Count part** | | | | | | | | | |
| Geographical distance (thous mi) | -0.342*** | -0.293*** | -0.223*** | -0.247*** | -0.269*** | -0.271*** | -0.196*** | -0.225*** | -0.248*** |
| Geographical distance squared (thous mi) | 0.016*** | 0.012*** | 0.008** | 0.010*** | 0.012*** | 0.010*** | 0.006* | 0.008** | 0.010*** |
| Number of papers at destination | 0.129*** | 0.124*** | 0.119*** | 0.117*** | 0.117*** | 0.117*** | 0.110*** | 0.109*** | 0.108*** |
| Number of papers at destination squared | -0.001*** | -0.001*** | -0.001*** | -0.001*** | -0.001*** | -0.001*** | -0.000*** | -0.000*** | -0.000*** |
| lines0stop | | 0.309*** | | | | 0.342*** | | | |
| lines0stop squared | | -0.024*** | | | | -0.026*** | | | |
| lines1stop | | | 0.075*** | | | | 0.079*** | | |
| lines1stop squared | | | -0.001*** | | | | -0.001*** | | |
| lines2stop | | | | 0.030*** | | | | 0.030*** | |
| lines2stop squared | | | | -0.000*** | | | | -0.000*** | |
| lines3stop | | | | | 0.005*** | | | | 0.005*** |
| lines3stop squared | | | | | -0.000*** | | | | -0.000*** |
| Distance to airport at destination (mi) | | | | | | -0.012*** | -0.013*** | -0.013*** | -0.013*** |
| Constant | 2.052*** | 1.918*** | 1.528*** | 1.371*** | 1.336*** | 2.154*** | 1.756*** | 1.606*** | 1.568*** |
| **Inflate part** | | | | | | | | | |
| Number of papers at destination | -3.787*** | -3.709*** | -3.681*** | -3.679*** | -3.678*** | -3.487*** | -3.438*** | -3.436*** | -3.438*** |
| Constant | -0.104 | -0.1 | -0.125 | -0.144* | -0.149* | -0.193** | -0.224** | -0.242** | -0.249*** |
| Constant lnalpha | 0.827*** | 0.814*** | 0.796*** | 0.796*** | 0.794*** | 0.796*** | 0.773*** | 0.773*** | 0.771*** |
| **Statistics** | | | | | | | | | |
| Observations | 8925 | 8925 | 8925 | 8925 | 8925 | 8925 | 8925 | 8925 | 8925 |
| AIC | 40998.1 | 40948.3 | 40813.1 | 40789.8 | 40773.8 | 40785.7 | 40626.8 | 40608.1 | 40590.9 |
| BIC | 41054.9 | 41019.3 | 40884.0 | 40860.7 | 40844.7 | 40863.8 | 40704.9 | 40686.2 | 40668.9 |

Significance levels: * p<0.05; ** p<0.01; *** p<0.001.

Not only the existence of flight connection matters, but also its passenger capacity. Taking into account the number of available seats improves model's fit as measured by AIC and BIC. This is visible by comparing models based on simple connectivity variable, 'LinesXstop' (Table 4), and models based on seats-weighted connectivity variable, 'SeatsXstop' (Table 5). In the case of specifications with direct connections (models 6 and 10), connections up to one stop (models 7 and 11), and connections up to two stops (models 8 and 12), BIC and AIC statistics are in favor of seats-weighted connectivity variable. However, in the case of connections up to three stops, non-weighted connectivity variable does better. This is probably because connections requiring up to three changes are rare, so in their case, the most important thing is the existence of a connection, not its capacity. Regardless, in the group of models presented in tables 4 and 5, model (12), involving seats-weighted connections up to two stops, has the lowest AIC and BIC values, and therefore it can be preferred as best suited to the analyzed data.

**Table 5. Research collaboration and air transport – seats capacity**

| Dependent variable: Number of co-authored papers | (10) | (11) | (12) | (13) |
|---|---|---|---|---|
| **Count part** | | | | |
| Geographical distance (thous mi) | -0.271*** | -0.242*** | -0.266*** | -0.280*** |
| Geographical distance squared (thous mi) | 0.010*** | 0.008** | 0.009*** | 0.011*** |
| Number of papers at destination | 0.116*** | 0.110*** | 0.109*** | 0.112*** |
| Number of papers at destination squared | -0.001*** | -0.000*** | -0.000*** | -0.000*** |
| Seats0stop | 2.665*** | | | |
| Seats0stop squared | -1.594*** | | | |
| Seats1stop | | 0.468*** | | |
| Seats1stop squared | | -0.044*** | | |
| Seats2stop | | | 0.184*** | |
| Seats2stop squared | | | -0.006*** | |
| Seats3stop | | | | 0.003*** |
| Seats3stop squared | | | | -0.000*** |
| Distance to airport at destination (mi) | -0.012*** | -0.013*** | -0.013*** | -0.013*** |
| Constant | 2.160*** | 1.897*** | 1.739*** | 1.957*** |
| **Inflate part** | | | | |
| Number of papers at destination | -3.479*** | -3.449*** | -3.441*** | -3.515*** |
| Constant | -0.191** | -0.225** | -0.248*** | -0.235** |
| Constant lnalpha | 0.794*** | 0.772*** | 0.769*** | 0.780*** |
| **Statistics** | | | | |
| Observations | 8925 | 8925 | 8925 | 8925 |
| AIC | 40782.5 | 40617.0 | 40578.8 | 40637.5 |
| BIC | 40860.5 | 40695.1 | 40656.9 | 40715.6 |

To ensure meaningful coefficients SeatsXstop variable is divided by 1000.
Significance levels: * $p<0.05$; ** $p<0.01$; *** $p<0.001$.

Further analysis of the compared models reveals, firstly, that direct connections have a stronger impact on the probability of scientific cooperation than flights requiring transfers—see specifications (14)-(18) with dummy variables for direct and connecting flights presented in Table 6. In the case of destinations that have no direct flight connection and requires minimum one stop, the number of expected co-publication decreases by a factor of 0.49 as compared to destinations that can be reached with a single flight (for a full dataset as specified by model 14). Secondly, the greater the number of transfers required, the weaker the effect on the dependent variable. This is evidenced by the fact that the models with only direct flights—specifications (2), (6), and (10)—have the highest coefficient of air transport variable (Lines0stop and Seats0Stop). In turn, models with up to one, two or three stops show decreasing values of air transport coefficient (Lines1stop and Seats1stop, Lines2stop and Seats2stop, Lines3stop and Seats3stop, respectively). This result is in line with expectations. Direct flights and those requiring fewer transfers are more convenient for passengers than connections requiring many stops. At the same time, not only air transport connectivity matters but also the distance between the location of the co-authors and their nearest airport. The results of the estimation confirm the common sense of expectations that the proximity of the airport is advantageous, at least in the case of long-distance cooperation, which from time to time requires air travel.

**Table 6. Research collaboration and air transport – direct and connecting flights**

| Dependent variable: Number of co-authored papers | **Full dataset** (14) | **ASU** (15) | **IUB** (16) | **IUPUI** (17) | **UMICH** (18) |
|---|---|---|---|---|---|
| **Count part** | | | | | |
| Geographical distance (thous mi) | -0.122*** | -0.420*** | -0.284*** | -0.374*** | -0.077 |
| Geographical distance squared (thous mi) | -0.000 | 0.022*** | 0.014* | 0.025*** | -0.005 |
| Number of papers at destination | 0.113*** | 0.108*** | 0.108*** | 0.099*** | 0.142*** |
| Number of papers at destination squared | -0.000*** | -0.000*** | -0.000*** | -0.000*** | -0.001*** |
| Minimum number of stops to reach destination (compared to direct flight): | | | | | |
| 1 stop | -0.705*** | -0.313* | -0.112 | -1.117*** | -0.399*** |
| 2 stops | -1.274*** | -0.340* | -0.575* | -1.275*** | -0.646*** |
| 3 stops | -1.617*** | -1.328*** | -0.824* | -1.237*** | -0.54 |
| Distance to airport at destination (mi) | -0.012*** | -0.011*** | -0.009*** | -0.011*** | -0.014*** |
| Constant | 2.743*** | 2.723*** | 2.118*** | 3.036*** | 2.622*** |
| **Inflate part** | | | | | |
| Number of papers at destination | -3.528*** | -3.748*** | -4.178*** | -1.666*** | 0.027 |
| Constant | -0.225** | 0.082 | 0.669*** | 0.364** | -23.779 |
| Constant lnalpha | 0.766*** | 0.553*** | 0.656*** | 0.664*** | 0.684*** |
| **Statistics** | | | | | |
| Observations | 8907 | 2241 | 2222 | 2223 | 2221 |
| AIC | 40522.8 | 9495.3 | 8475.1 | 8135.1 | 13538.4 |
| BIC | 40607.9 | 9563.9 | 8543.6 | 8203.5 | 13606.8 |

Significance levels: * $p<0.05$; ** $p<0.01$; *** $p<0.001$.

The relationship between air connectivity and the number of co-authored papers is not linear. All the squared air connectivity variables are significant in specifications (1)-(13). Negative coefficients of the quadratic terms suggest that at some point, the connectivity is so high that its further increase (e.g.

adding one more flight between given airports) has far less impact on collaboration than the similar increase at low levels of the overall connectivity.

The number of scientific papers affiliated in potentially cooperating destinations serves two functions in presented models: first, as specified in the inflate part, and second, as specified in the count part. The count part can be interpreted similarly to standard maximum likelihood models. Firstly, the increase in the number of articles at destination translates into reduction in the likelihood of a complete absence of co-authored articles. In other words, an increase in the number of articles at destination decreases the likelihood that the variable 'number of co-authored articles' will equal zero. Secondly, as the count part of the models shows, the more articles in the cooperating destination, the higher the number of co-authored papers between the ego and the destination. However, this relationship is more complex, as indicated by the significant quadratic term for the number of articles at the destination. Negative coefficients of the quadratic term indicate the curvilinear shape of the relationship: as the number of articles increases, its positive influence on the number of co-authored articles is flattening out.

In all presented models, geographical distance is negatively associated with research collaboration. The higher the distance, the smaller the number of co-publications. Furthermore, the effect is also curvilinear. In this case, positive coefficient of the squared variable suggests that the negative influence of physical distance on collaboration decreases gradually as the geographic separation increases. This can be interpreted as follows: the difference between, for example, 9,100 or 9,200 miles does not translate into a significant difference for the person considering a trip to such a remote place. But the difference between 100 and 200 miles means, approximately, a two-fold lengthening of the journey and thus, can be a significant factor influencing the decision.

The influence of geographical distance on the number of co-publications is modified by air transport connectivity and accessibility, as well as by scientific capacity of collaborators. The low number of papers at the destination, less than one thousand, usually translates into the low number of co-publications, no matter the distance. On the other hand, for destinations that accumulated high research capacity, the distance matters a lot. For example, in the case of destinations with 30 thousand papers, the decrease in the distance from 4,000 to 1,000 miles raises the expected number of co-publications twice, from circa 50 to 100. While the decrease from 10,000 to 7,000 miles (i.e., by the same number of miles, 3,000), raises the expected number of co-publications by no more than ten papers. Similarly, for the low values of connectivity and accessibility, the relation between geographical distance and expected number of co-authrships is flatter than for high values of those variables. Furthermore, the distance matters significantly more in the case of direct flights, than for connections requiring one, and in particular, two or three transfers (see Fig 3). This is reasonable as direct flights are constrained by technical capacities of aircrafts, as well as regulatory requirements, in particular limits for flight duty periods for crew member's (Campante & Yanagizawa-Drott, 2017).

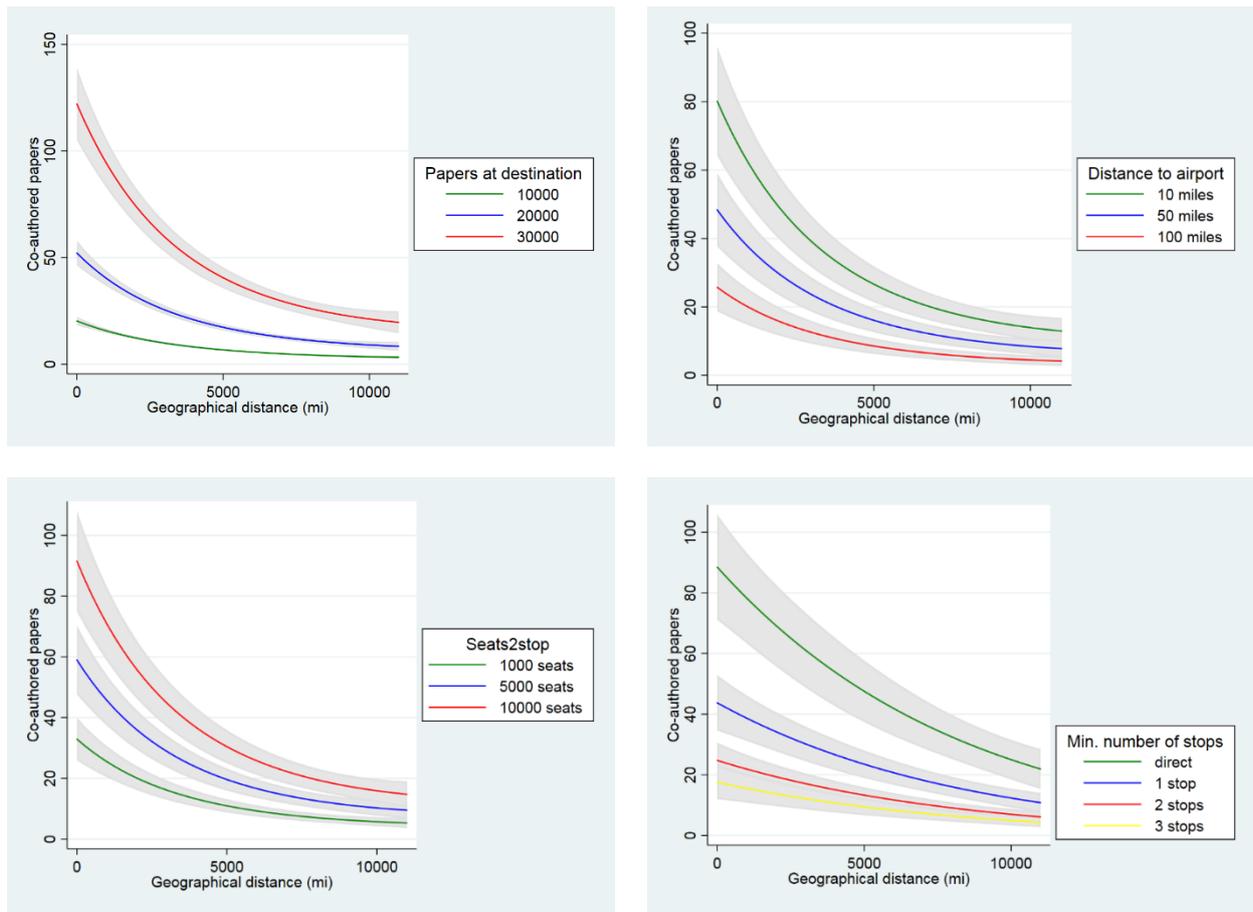

**Fig 3. Predicted number of collaborative papers at different values of selected independent variables***

* The estimations are based on model (12) in the case of 'Papers at destination', Searts2stop', and 'Distance to airport'. For 'Minimum number of stops' model 14 has been employed.

Estimations based on institutional sub-datasets (Table 6 and 7) show that relationship between air transport connectivity and research collaboration is not homogeneous across universities. Comparison of IUB and IUPUI is particularly interesting. Both institutions are served by the same airport. Thus they have the same air transport connectivity (However, it should be emphasized that IUB is located at a much greater distance to the Indianapolis airport than IUPUI). In the case of IUPUI, direct flights are the most significant predictors of co-publications, both statistically and substantially. While for IUB the availability of direct flights is not essential, but connections up to one and two stops matters much more than for IUPUI—compare specifications (23)-(30). Such divergent patterns can be attributed to organization-specific research collaboration networks, related to the disciplinary composition of institutions (IUPUI hosts school of medicine, while IUB does not), and resulted from long-term path-dependent processes.

**Table 7. Research collaboration and air transport—institutional sub-datasets**

| Dependent variable: Number of co-authored papers | ASU | | | | IUB | | | | IUPUI | | | | UMICH | | | |
|---|---|---|---|---|---|---|---|---|---|---|---|---|---|---|---|---|
| | (19) | (20) | (21) | (22) | (23) | (24) | (25) | (26) | (27) | (28) | (29) | (30) | (31) | (32) | (33) | (34) |
| **Count part** | | | | | | | | | | | | | | | | |
| Geographical distance (thous mi) | -0.432*** | -0.439*** | -0.427*** | -0.461*** | -0.406*** | -0.347*** | -0.378*** | -0.399*** | -0.434*** | -0.394*** | -0.421*** | -0.468*** | -0.141** | -0.166*** | -0.171*** | -0.184*** |
| Geographical distance squared (thous mi) | 0.023*** | 0.024*** | 0.022*** | 0.025*** | 0.022*** | 0.018** | 0.020*** | 0.022*** | 0.031*** | 0.028*** | 0.030*** | 0.034*** | -0.001 | 0.001 | 0.001 | 0.003 |
| Number of papers at destination | 0.108*** | 0.106*** | 0.104*** | 0.108*** | 0.108*** | 0.105*** | 0.102*** | 0.105*** | 0.098*** | 0.096*** | 0.095*** | 0.100*** | 0.145*** | 0.134*** | 0.134*** | 0.136*** |
| Number of papers at destination squared | -0.000*** | -0.000*** | -0.000*** | -0.000*** | -0.000*** | -0.000*** | -0.000*** | -0.000*** | -0.000*** | -0.000*** | -0.000*** | -0.000*** | -0.001*** | -0.001*** | -0.001*** | -0.001*** |
| Seats0stop | 1.196* | | | | 1.112 | | | | 6.021*** | | | | 1.805* | | | |
| Seats0stop squared | -0.572 | | | | -1.245 | | | | -5.683** | | | | -1.483* | | | |
| Seats1stop | | 0.128* | | | | 0.433*** | | | | 0.308* | | | | 0.296*** | | |
| Seats1stop squared | | -0.008 | | | | -0.080* | | | | -0.008 | | | | -0.029*** | | |
| Seats2stop | | | 0.091*** | | | | 0.161*** | | | | 0.122** | | | | 0.130*** | |
| Seats2stop squared | | | -0.003* | | | | -0.009* | | | | -0.003 | | | | -0.005*** | |
| Seats3stop | | | | 0.001 | | | | 0.002** | | | | 0.001 | | | | 0.002*** |
| Seats3stop squared | | | | 0 | | | | -0.000* | | | | 0 | | | | -0.000** |
| Distance to airport at destination (mi) | -0.012*** | -0.012*** | -0.012*** | -0.012*** | -0.009*** | -0.010*** | -0.010*** | -0.010*** | -0.010*** | -0.010*** | -0.010*** | -0.010*** | -0.014*** | -0.014*** | -0.014*** | -0.014*** |
| Constant | 2.426*** | 2.396*** | 2.220*** | 2.461*** | 2.030*** | 1.789*** | 1.711*** | 1.942*** | 1.944*** | 1.731*** | 1.663*** | 1.957*** | 2.311*** | 2.277*** | 2.116*** | 2.320*** |
| **Inflate part** | | | | | | | | | | | | | | | | |
| Number of papers at destination | -3.735*** | -3.720*** | -3.698*** | -3.738*** | -4.164*** | -4.141*** | -4.112*** | -4.135*** | -1.631*** | -1.664*** | -1.653*** | -1.715*** | 0.032 | -5.637 | -5.573 | -5.634 |
| Constant | 0.098 | 0.09 | 0.073 | 0.093 | 0.697*** | 0.670*** | 0.658*** | 0.673*** | 0.382** | 0.355** | 0.339** | 0.359** | -19.984 | -1.806*** | -1.823*** | -1.809*** |
| Constant lnalpha | 0.554*** | 0.556*** | 0.552*** | 0.559*** | 0.665*** | 0.659*** | 0.655*** | 0.662*** | 0.664*** | 0.679*** | 0.682*** | 0.698*** | 0.691*** | 0.599*** | 0.594*** | 0.605*** |
| **Statistics** | | | | | | | | | | | | | | | | |
| Observations | 2244 | 2244 | 2244 | 2244 | 2228 | 2228 | 2228 | 2228 | 2229 | 2229 | 2229 | 2229 | 2224 | 2224 | 2224 | 2224 |
| AIC | 9506.6 | 9506.5 | 9497.2 | 9510.4 | 8508.4 | 8491.1 | 8485.9 | 8496.7 | 8155.1 | 8158.9 | 8158.1 | 8174.7 | 13566.3 | 13517.5 | 13505.9 | 13528.2 |
| BIC | 9569.5 | 9569.4 | 9560 | 9573.3 | 8571.2 | 8553.9 | 8548.7 | 8559.5 | 8217.9 | 8221.7 | 8220.9 | 8237.5 | 13629 | 13580.3 | 13568.7 | 13590.9 |

To ensure meaningful coefficients SeatsXstop variable is divided by 1000.
Significance levels: * $p<0.05$; ** $p<0.01$; *** $p<0.001$.

# DISCUSSION AND CONCLUSIONS

The paper makes two contributions. First, we show that air transport availability is an important factor for scientific collaboration, even when controlling for geographical distance and research capacities of collaborators. Second, both air transport connectivity (direct and indirect air connections between airports) and accessibility (distance to the nearest airport) are important correlates of scientific collaboration. Presented estimation results provide evidence that more flight connections and greater seat capacity increase the number of co-publications. Also, proximity of airport at collaborating destination is positively related to the expected number of co-authored papers. Moreover, direct flights and flights with one transfer are more valuable for intensifying scientific collaboration than travels involving more connecting flights. One additional direct flight rise the expected number of co-publications by a factor of 1.41, while additional connection requiring up to two stops rises the number by a factor of 1.03. The results of our study are in line with conclusions from broader research corpus highlighting the importance of air transport for the economic development of cities and regions (DSA et al., 2013). In particular, the availability of direct flights is seen as a significant predictor of a city's fortunes (Campante & Yanagizawa-Drott, 2017).

Estimations based on four separate institutional sub-datasets show that the relationship between transport accessibility and scientific cooperation is not uniform. For some institutions—Indiana University-Purdue University Indianapolis in the first place—direct flights are more valuable predictors of distant co-publications, while for other three institutions indirect connections up to one or two stops better explain collaboration patterns. This diversity can be related to different research profiles of studied universities (see Supporting information). Not only research organizations differ in scientific specialization, but also scientific disciplines are spatially biased regarding propensity to collaborate (Wagner, 2008; Ponds et al., 2007). For example, collaboration in experimental particle physics is far more spatially bound than collaboration in theoretical mathematics. This organizational and disciplinary diversity shapes spatial patterns of collaboration, in a dynamic coopetitive processes (Nickelsen & Krämer, 2016).

Two limitations of the presented approach have to be underlined. First, the direction of the relationship between air transport availability and research collaboration is ambiguous. Increasing collaboration can be both the result and the cause of transport availability. Development of collaborative relations between distant locations indeed rises the demand for transport. However, based on the results of a quasi-experimental study by Catalini, Fons-Rosen & Gaulé (2016), we can expect that causal relation from transport connectivity to scientific collaboration also happen. Moreover, the circular cumulative causation can be expected—i.e., more collaboration leads to higher transport demand and in result greater transport capacity, which in turn induces more collaboration, and so forth. The second limitation is related to the dataset used in this study. We focused on four selected universities located in the US.

In other socio-economic and geographical contexts, the role of air transport can be different. For example, in Europe, Japan, and increasingly in China, railway connectivity can be more critical than air transport, at least up to some geographical distance.

Future studies should take into account overall geo-localized co-authorship network instead of looking on selected ego-networks. Second, controlling for more variables might deepen our understanding of the phenomena. In particular, citation data can improve research capacity measurement. Third, other modes and measures of research collaboration should be examined. Patent data are most promising as they are easily accessible. Four, different modes of transport should be incorporated in future studies, in particular road and railroad accessibility. We expect that essential insights can be gained by combining multimodal transport connectivity and multimodal research collaborations, comparing and integrating co-publications, co-patenting and collaborative research projects.

# ACKNOWLEDGEMENTS

We would like to thank Patty Mabry and Ann McCranie (Indiana University Bloomington) for comments on a previous version of the paper, as well as the participants of the 12th Workshop on the Organization, Economics and Policy of Scientific Research (University of bath, UK, 2018) for ideas to expand the scope of the study. Responsibility for all errors remains our own.

This work uses Web of Science data by Clarivate Analytics provided by the Network Science Institute and the Cyberinfrastructure for Network Science Center at Indiana University. Katy Börner is partially supported by the National Institutes of Health under awards P01AG039347 and U01CA198934 and National Science Foundation awards AISL1713567 and OAC1445604. Adam Ploszaj is partially supported by grant 2011/03/B/HS4/05737 from the Polish National Science Center. Any opinions, findings, and conclusions or recommendations expressed in this material are those of the author(s) and do not necessarily reflect the views of the National Science Foundation or the Polish National Science Center.

# REFERENCES

Adams J. Collaborations: The fourth age of research. Nature. 2013;497(7451): 557.

Allen JT. Managing the Flow of Technology: Technology Transfer and the Dissemination of Technologycal Information within the R&D Organization. Cambridge, MA: The Massachusetts Institute of Technology; 1977.

Andersson ÅE, Persson O. Networking scientists. The Annals of Regional Science. 1993;27(1): 11-21.

Andersson M, Ejermo O. How does accessibility to knowledge sources affect the innovativeness of corporations?—evidence from Sweden. The annals of regional science. 2005;39(4): 741-65.

Barber MJ, Scherngell T. Is the European R&D network homogeneous? Distinguishing relevant network communities using graph theoretic and spatial interaction modelling approaches. Regional Studies. 2013;47(8): 1283-98.

Bianconi G, Barabási AL. Competition and multiscaling in evolving networks. EPL (Europhysics Letters). 2001;54(4): 436.

Boschma R. Proximity and innovation: a critical assessment. Regional studies. 2005;39(1): 61-74.

Boudreau KJ, Brady T, Ganguli I, Gaule P, Guinan E, Hollenberg A, Lakhani KR. A field experiment on search costs and the formation of scientific collaborations. Review of Economics and Statistics. 2017;99(4): 565-76.

Burnham KP, Anderson DR. Model selection and multimodel inference: a practical information-theoretic approach. 2nd ed. New York: Springer; 2002.

Cairncross F. The Death of Distance: How the Communications Revolution Will Change Our Lives. Boston: Harvard Business School Press; 1997.

Campante F, Yanagizawa-Drott D. Long-range growth: economic development in the global network of air links. The Quarterly Journal of Economics. 2017;133(3): 1395-458.

Capello R, Caragliu A. Proximities and the intensity of scientific relations: synergies and nonlinearities. International Regional Science Review. 2018;41(1): 7-44.

Catalini C, Fons-Rosen C, Gaulé P. Did Cheaper Flights Change the Geography of Scientific Collaboration? MIT Sloan Research Paper. 2016 No. 5172-16.

Catalini C. Microgeography and the Direction of Inventive Activity. Management Science. 2018;64(9): 4348-4364.

Dong X, Zheng S, Kahn ME. The Role of Transportation Speed in Facilitating High Skilled Teamwork. National Bureau of Economic Research. 2018. Working Paper 24539.

DSA et al. ESPON ADES project: Airports as Drivers of Economic Success in Peripheral Regions, Final Report, 2013.

Ejermo O, Karlsson C. Interregional inventor networks as studied by patent coinventorships. Research Policy. 2006;35(3): 412-30.

Fernández A, Ferrándiz E, León MD. Proximity dimensions and scientific collaboration among academic institutions in Europe: The closer, the better?. Scientometrics. 2016;106(3): 1073-92.

Franceschet M, Costantini A. The effect of scholar collaboration on impact and quality of academic papers. Journal of informetrics. 2010;4(4): 540-53.

Frenken K, Hardeman S, Hoekman J. Spatial scientometrics: Towards a cumulative research program. Journal of Informetrics. 2009;3(3): 222-32.

Friedman TL. The world is flat: A brief history of the twenty-first century. New York: Farrar, Straus and Giroux; 2005.

Gingras Y. Bibliometrics and Research Evaluation. Uses and Abuses. Cambridge, Massachusetts: The MIT Press; 2016.

Hoekman J, Frenken K, Tijssen RJ. Research collaboration at a distance: Changing spatial patterns of scientific collaboration within Europe. Research Policy. 2010;39(5): 662-73.

Hoekman J, Frenken K, Van Oort F. The geography of collaborative knowledge production in Europe. The Annals of Regional Science. 2009;43(3): 721-38.

Hoekman J, Scherngell T, Frenken K, Tijssen R. Acquisition of European research funds and its effect on international scientific collaboration. Journal of Economic Geography. 2013;13(1): 23-52.

Hua CI, Porell F. A critical review of the development of the gravity model. International Regional Science Review. 1979;4(2): 97-126.

Kabo F, Hwang Y, Levenstein M, Owen-Smith J. Shared paths to the lab: A sociospatial network analysis of collaboration. Environment and Behavior. 2015;47(1): 57-84.

Kabo FW, Cotton-Nessler N, Hwang Y, Levenstein MC, Owen-Smith J. Proximity effects on the dynamics and outcomes of scientific collaborations. Research Policy. 2014 Nov 1;43(9): 1469-85.

Katz JS. Geographical proximity and scientific collaboration. Scientometrics. 1994;31(1): 31-43.

Ke W. A fitness model for scholarly impact analysis. Scientometrics. 2013;94(3): 981-98.

Knoben J, Oerlemans LA. Proximity and inter-organizational collaboration: A literature review. International Journal of Management Reviews. 2006;8(2): 71-89.

Larivière V, Gingras Y, Archambault É. Canadian collaboration networks: A comparative analysis of the natural sciences, social sciences and the humanities. Scientometrics. 2006;68(3): 519-33.

Lata R, Scherngell T, Brenner T. Integration Processes in European Research and Development: A Comparative Spatial Interaction Approach Using Project Based Research and Development Networks, Co-Patent Networks and Co-Publication Networks. Geographical Analysis. 2015;47(4): 349-75.

Long JS. Regression models for categorical and limited dependent variables. Advanced Quantitative Techniques in the Social Sciences Series, vol. 7. Beverly Hills, CA: Sage; 1997.

Ma H, Fang C, Pang B, Li G. The effect of geographical proximity on scientific cooperation among Chinese cities from 1990 to 2010. PLOS ONE. 2014;9(11): e111705.

Maisonobe M, Eckert D, Grossetti M, Jégou L, Milard B. The world network of scientific collaborations between cities: domestic or international dynamics?. Journal of Informetrics. 2016;10(4): 1025-36.

Marek P, Titze M, Fuhrmeister C, Blum U. R&D collaborations and the role of proximity. Regional Studies. 2017;51(12): 1761-73.

Matthiessen CW, Schwarz AW, Find S. World cities of scientific knowledge: Systems, networks and potential dynamics. An analysis based on bibliometric indicators. Urban Studies. 2010;47(9): 1879-97.

Mazloumian A, Helbing D, Lozano S, Light RP, Börner K. Global multi-level analysis of the 'Scientific Food Web'. Scientific reports. 2013;3: 1167.

Morgan K. The exaggerated death of geography: learning, proximity and territorial innovation systems. Journal of economic geography. 2004;4(1): 3-21.

Nagpaul P. Exploring a pseudo-regression model of transnational cooperation in science. Scientometrics. 2003;56(3): 403-16.

Nickelsen K, Krämer F. Introduction: Cooperation and Competition in the Sciences. NTM Zeitschrift für Geschichte der Wissenschaften, Technik und Medizin. 2016;24(2): 119-123.

Olechnicka A, Ploszaj A, Celinska-Janowicz D. The Geography of Scientific Collaboration. Abingdon & New York: Routledge; 2018.

Olson GM, Olson JS. Distance matters. Human–computer interaction. 2000;15(2-3): 139-78.

Perianes-Rodriguez A, Waltman L, van Eck NJ. Constructing bibliometric networks: A comparison between full and fractional counting. Journal of Informetrics. 2016;10(4): 1178-95.

Picci L. The internationalization of inventive activity: A gravity model using patent data. Research Policy. 2010;39(8): 1070-81.

Plotnikova T, Rake B. Collaboration in pharmaceutical research: exploration of country-level determinants. Scientometrics. 2014;98(2): 1173-202.

Ponds R, Van Oort F, Frenken K. The geographical and institutional proximity of research collaboration. Papers in regional science. 2007;86(3): 423-43.

Raftery AE. Bayesian model selection in social research. Sociological methodology. 1995;25: 111-63.

Sebestyén T, Varga A. Research productivity and the quality of interregional knowledge networks. The Annals of Regional Science. 2013;51(1): 155-89.

Wagner C, Park H, Leydesdorff L. The Continuing Growth of Global Cooperation Networks in Research: A Conundrum for National Governments. PLOS ONE. 2015;10(7): e0131816.

Wagner CS. The new invisible college: science for development. Washington, D.C.: Brookings Institution Press; 2008.

Waltman L, Tijssen RJ, van Eck NJ. Globalisation of science in kilometres. Journal of Informetrics. 2011;5(4): 574-82.

Zitt M, Ramanana-Rahary S, Bassecoulard E. Correcting glasses help fair comparisons in international science landscape: Country indicators as a function of ISI database delineation. Scientometrics. 2003;56(2): 259-82.